\documentstyle[aps]{revtex}
\input epsf
\def\jepsfbox#1{\typeout{#1} \epsfbox{#1}}
\def\plotonesc#1#2{\begin{center} \leavevmode
\epsfxsize=#2\columnwidth \jepsfbox{#1} \end{center}}
\def\plottwo#1#2{\centering \leavevmode
\epsfxsize=.45\columnwidth \jepsfbox{#1} \hfil
\epsfxsize=.45\columnwidth \jepsfbox{#2}}
\def\jcite#1#2{#1 \cite{#2}}
\def\rarrow{\rightarrow}
\def\etal{{\it et al.\ }}
\def\eg{{\it e.g.~}}
\def\ie{{\it i.e.~}}

\def\rmmat#1{{\hbox{\rm #1}}}
\def\rmscr#1{\rmmat{\scriptsize #1}}
\newcommand{\lel}{{\lambda_e^{\!\!\!\!-}}}
\newcommand{\be}{\begin{equation}}
\newcommand{\ee}{\end{equation}}
\newcommand{\ba}{\begin{eqnarray}}
\newcommand{\ea}{\end{eqnarray}}
\def\p{\partial}
\def\d{{\rm d}}

\def\pp#1#2{\frac{\p #1}{\p #2}}
\def\figref#1{Fig.~\ref{fig:#1}}
\def\tabref#1{Table~\ref{tab:#1}}
\def\eqref#1{Eq.~\ref{eq:#1}}
\begin{document}
\draft
\newcommand{\bfi}{{\bf B}} \newcommand{\efi}{{\bf E}}
\newcommand{\lag}{{\cal L}} \newcommand{\dLIII}{{\frac{\partial^3
\lag}{\partial I^3}}} \newcommand{\dLII}{{\frac{\partial^2
\lag}{\partial I^2}}} \newcommand{\dLI}{{\frac{\partial \lag}{\partial
I}}} \newcommand{\dLKKK}{{\frac{\partial^3 \lag}{\partial K^3}}}
\newcommand{\dLKK}{{\frac{\partial^2 \lag}{\partial K^2}}}
\newcommand{\dLK}{{\frac{\partial \lag}{\partial K}}}
\newcommand{\dLIK}{{\frac{\partial^2 \lag}{\partial I \partial K}}}
\title{Hydrogen and Helium Atoms and Molecules in an Intense 
Magnetic Field}
\author{Jeremy S. Heyl\thanks{Current Address: Theoretical Astrophysics, mail code 130-33,
California Institute of Technology, Pasadena, CA 91125}
\and Lars Hernquist\thanks{Presidential Faculty
Fellow}}
\address{Lick Observatory,
University of California, Santa Cruz, California 95064, USA}
\maketitle
\begin{abstract}
We calculate the atomic structure of hydrogen and helium, atoms and
molecules in an intense magnetic field, analytically and
numerically with a judiciously chosen basis.
\end{abstract}
\pacs{31.10.+z, 31.15.-p, 32.60.+i, 97.60.Jd}
\section{Introduction}

The problem of atoms and molecules in a magnetic field is both a
classic example of time-independent perturbation theory and a vexing
challenge in the study of neutron stars and white dwarfs.  A
sufficiently intense magnetic field cannot be treated perturbatively.
The spectra and properties of neutron-star atmospheres depend crucially
on magnetic field.  Indeed, in the intense magnetic field of a neutron
star $B\gtrsim 10^{10}$ G the nucleus rather than the field acts as a
perturbation.  The electron is effectively confined to move along the
magnetic field lines.

This work extends classic analytic work on the one-dimensional hydrogen
atom \cite{Loud59,Hain69} to form the basis of a perturbative treatment
of hydrogen in an intense magnetic field.  This analytic treatment
yields binding energies for $B\gtrsim 10^{12}$ G whose accuracy rivals
that of the recent exhaustive treatment of hydrogen in an magnetic field
by \jcite{Ruder \etal}{Rude94} with substantially less computational
expense.

We also present a straightforward numerical treatment of the hydrogen
atom, the hydrogen molecular ion and the helium atom.  The electron
wavefunction is expanded in a convenient basis, and the Schr{\"o}dinger
equation may be solved approximately by diagonalizing a suitably
calculated matrix.  The effective potential between the electrons and
between the electrons and the nuclei may be determined analytically,
expediting the calculation dramatically.

\section{The Single Electron Problem}
\def\nablab{\hbox{\boldmath $\nabla$}}
We begin with the problem of a single electron bound by the combined
field of an atomic nucleus and strong external magnetic field.  The
Hamiltonian for the electron is given by
\be
H = \frac{{\bf P}^2}{2 M} - \frac{Z e^2}{r} 
- \mu \cdot {\bf B}
\ee
where we have assumed that the nucleus is infinitely massive,  $M$ is
the mass of the electron and ${\bf P} = {\bf p} - e/c\,{\bf A}$.  

To derive the Schr\"{o}dinger equation for the electron, we make the
replacement ${\bf p}=-i \hbar \nablab$.  We take the magnetic
field to point in the $z$-direction and choose the gauge where
$A_\phi=B \rho/2$, $A_\rho=A_z=0$ and obtain
\be
\left ( -\frac{\hbar^2}{2 M} \nablab^2 
- \frac{i \hbar}{2 M c} B |e| \pp{}{\phi} 
+ \frac{1}{8} \frac{e^2}{M c^2} B^2 \rho^2 - \frac{Z e^2}{r}
 - \frac{\mu}{s} \sigma_z B - E \right ) \psi(1) = 0
\ee
where $1$ denotes the spin and spatial coordinates of the electron 
\ie ${\bf r}_1, \sigma_1$.  The spin portion of the wavefunction
decouples from the spatial component; therefore, we take the electron
spins antialigned with the magnetic field to minimize the total
energy, \ie to calculate the ground state.

For $Z=0$, we recover the equation for a free electron in an external
magnetic field which is satisfied by a function of the form
\be
\psi_{n m p_z} ({\bf r}) =
 R_{n m}(\rho,\phi) e^{i z p_z/\hbar}
\ee
where
\def\F11{\mbox{ }_1\! F_1}
\be
R_{n m}(\rho,\phi) = \frac{1}{a_H^{|m|+1} |m|!} \left [ 
\frac{(|m|+n)!}{2^{|m|+1} \pi n!} \right ]^{1/2} 
\exp \left ( - \frac{\rho^2}{4 a_H^2} \right ) \rho^{|m|}
\F11(-n,|m|+1,\rho^2/2 a_H^2) e^{i m \phi},
\ee
where $a_H=\sqrt{\hbar/M \omega_H}=\sqrt{\hbar c/|e| B}$ \cite{Land3},
and $\F11$ is the confluent hypergeometric function.  

It is convenient to define a critical field where the energy of the
Landau ground state $\hbar \omega_H/2$ equals the characteristic
energy of hydrogen $e^2/a_0$, where the Bohr radius, $a_0 \approx 
0.53 \AA$. The transition to the intense magnetic field regime (IMF)
occurs at \cite{Canu72}
\be
B_I = 2 m^2 c \left ( \frac{e}{\hbar} \right )^3 \approx 4.701 \times 10^9 
\rmmat{~G}.
\ee
We will express field strengths in terms of $\beta=B/B_I$.

For $Z\neq 0$, the complete solution may be expanded in a sum of
$\psi_{n m p_z}$ since these form a complete set.  However, for
sufficiently strong fields, one can treat the Coulomb potential as a
perturbation and use the ground Landau state with the appropriate $m$
quantum number as the first approximation to the radial wavefunction;
this is known as the adiabatic approximation.  

Equivalently, the adiabatic approximation assumes that the Coulomb
potential does not effectively
 mix the Landau states, \ie
\be
\left | \frac{ \langle R_{nm}|V(r)|R_{n'm} \rangle }{E_n-E_n'} \right | \ll\ 1.
\ee
To determine the validity of the adiabatic approximation we calculate this 
quantity for the first two Landau states and $m=0$,
\ba
\left | \frac{ \langle R_{00}|V(r)|R_{10} \rangle }{2 \alpha^2 \beta M c^2} \right | 
&=& \left | \frac{1}{2 \alpha^2 \beta M c^2} \frac{1}{2 \pi a_H^2} 
\int_0^\infty - \frac{Z e^2}{\sqrt{z^2+\rho^2}} \left ( 1 -
\frac{\rho^2}{2 a_H^2} \right ) 
\exp \left ( - \frac{\rho^2}{2 a_H^2} \right ) 2 \pi \rho \d \rho
\right | \\
&\leq& \left | \frac{1}{2 \alpha^2 \beta M c^2} \frac{1}{a_H^2} 
\int_0^\infty -Z e^2 
\left ( 1 - \frac{\rho^2}{2 a_H^2} \right ) 
\exp \left ( - \frac{\rho^2}{2 a_H^2} \right ) \d \rho \right | 
= \frac{Z}{4} \sqrt{\frac{\pi}{\beta}}.
\ea
where $\alpha\approx 1/137$ is the fine structure constant. 
We find for $\beta=1000$ ($B=4.7 \times 10^{12}$ G), that the
Coulomb potential mixes the Landau states of hydrogen by at most 
1.4 \%.  For stronger fields, the mixing is even less important.

In the adiabatic approximation, we assume that
\be
\psi_{0 m \nu}(1) = R_{0m}(\rho,\phi) Z_{m \nu} (z) \chi(\sigma)
\ee
where $Z_{m \nu}(z)$ remains to be determined, $\nu$ counts the number
of nodes in the $z$ wavefunction, and we expect the axial wavefunctions
to be different for different values of the magnetic quantum number
$m$.  We will use the notation, $|0 m \nu \rangle$, to designate the 
eigenstates.  

For $n=0$, the functions $R_{n m}$ assume a simple form
\be
R_{0m} (\rho,\phi) = {1 \over \sqrt{2^{|m|+1} \pi |m|!} a_H^{|m|+1}}
	\rho^{|m|} \exp \left ( -
	{\rho^2 \over 4 a_H^2} \right ) e^{i m \phi}
\ee
\be
\left | R_{0m} (\rho,\phi) \right |^2 =
{(-1)^{|m|} \over 2 \pi |m|! } { 1 \over a_H^2} 
\left . \left ( \d \over \d \kappa \right )^{|m|} \left [
\exp \left ( - \kappa {\rho^2 \over 2 a_H^2} \right ) \right ]
\right |_{\kappa=1}
\ee
With these assumptions the functions $Z_{\nu m}(z)$ satisfy a one-dimensional
Schr\"{o}dinger equation,
\be
(H_z - E) Z = \left [ -\frac{\hbar^2}{2 M} \frac{\d^2}{\d z^2} +
V_{\rmscr{eff},0m}(z) - E_{\nu m} \right ] Z_{\nu m}(z) = 0,
\label{eq:zSchro}
\ee
where
\be
V_{\rmscr{eff},0m}(z) = \langle R|V(r)|R \rangle  = \int_0^\infty -{Z e^2 \over \sqrt{z^2+\rho^2}}
\left | R_{0m}(\rho) \right |^2 2 \pi \rho\,\d \rho.
\label{eq:Veff}
\ee
Performing the integral yields \cite{Lai92,Canu72}
\be
V_{\rmscr{eff},0m}(z) = -{Z e^2 \over a_H} \sqrt{\pi/2} {(-1)^{|m|}\over |m|! } \left. \left ( \d \over \d \kappa
\right )^{|m|} \left [ {1 \over \sqrt{\kappa}} \exp(\kappa z^2/2 a_H^2) 
\rmmat{erfc}(\sqrt{\kappa} |z|/\sqrt{2} a_H) \right ] \right |_{\kappa=1}
\label{eq:pot}
\ee
which for large $z$ approaches  $-Z e^2/z$.  The Schr\"odinger equation
with this potential is not tractable analytically.  We can take one of
two paths.  First, the potential may be replaced by a simpler yet still
accurate function.  Second, we attempt to solve the Schr\"odinger
equation numerically.

\section{Analytic Solution}

\subsection{The Approximate Potential
}
The potential given in \eqref{pot} for arbitrary $m$ may be approximated to
within 30 \% over the entire domain by the much simpler form
\be
V_{\rmscr{eff},0m}(z) \approx V_{\rmscr{approx},0m}(z) =
 - \frac{Z e^2}{|z|+k_m a_H} 
\label{eq:potapprox}
\ee
where 
\be
k_m = - \frac{Z e^2}{a_H V_{\rmscr{eff},0m}(0)} = 
 \sqrt{2} \frac{\Gamma(|m|+1)}{\Gamma(|m|+\frac{1}{2})} 
= \sqrt{\frac{2}{\pi}} \frac{2^{|m|} |m|!}{(2|m|-1)!!} 
\ee
The double factorial is defined by $(-1)!!=1$ and
$(2n+1)!!=(2n+1)(2n-1)!!$.  For large $m$, $\frac{1}{2} k_m a_H$
 asymptotically approaches $\sqrt{2|m|+1} a_H$, the mean radius of a 
Landau orbital. 

\begin{figure}
\plotonesc{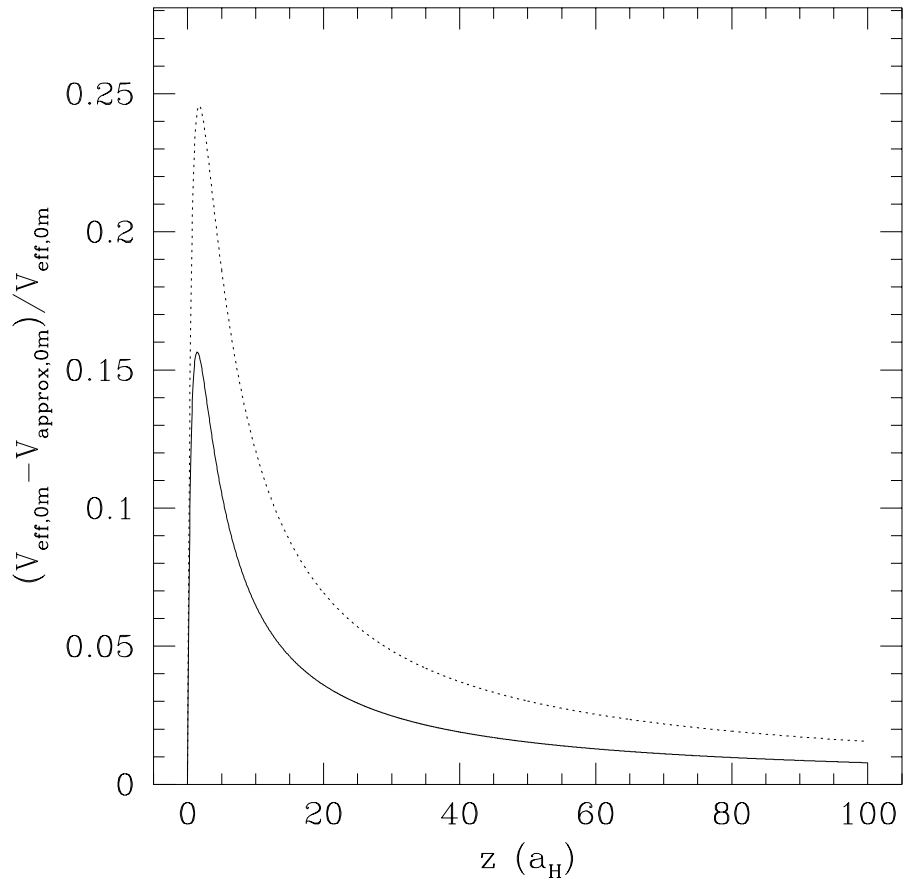}{0.7}
\caption[{The relative difference between the effective potential and the approximated potential}]
{The relative difference between the effective potential and the approximated potential.  The solid line
traces the difference for the $m=0$ state and the dotted line gives the $m=-1$ state.}
\label{fig:potcmp}
\end{figure}

As we see from \figref{potcmp}, the relative difference between the two
expressions is largest near $z=k_m a_H$. For $m=0$, the difference is
greater than 5 \% from $z=0.1$ to $z=10$.  We do not expect this
approximation to yield eigenvalues accurate to better than $\sim 10 \%$
for wavefunctions peaked in this range. 

We obtain the following eigenvalue equation with the approximated
potential,
\be
\left [ -\frac{\hbar^2}{2 M} \frac{\d^2}{\d z^2} -
\frac{Z e^2}{|z|+k_m a_H} - E_{0 m \nu} \right ] Z(z) = 0.
\label{eq:Schapprox}
\ee 
This equation is nearly identical to the
Schr\"odinger equation with a Coulomb potential; therefore, we treat it
as a Coulomb problem by using the natural units (Bohr
radii for length and Rydbergs for energy),
\be
z = \frac{\lel}{\alpha} \zeta \rmmat {~and~} \epsilon=\frac{2 E}{\alpha^2 M c^2}
\ee
which yields
\be
\left [ 
\frac{\d^2}{\d \zeta^2} +
 \left ( \epsilon + 
 \frac{2 Z}{|\zeta|+\zeta_m}
 \right ) \right ] Z(\zeta) = 0.
\ee
where $\zeta_m = k_m \sqrt{1/2\beta}$

Again as in the Coulomb problem, we perform the following substitutions
\be
n=\frac{1}{\sqrt{-\epsilon}} \rmmat{~and~} \xi=\frac{2\zeta}{n},
\ee
yielding,
\be
\left [ \frac{\d^2}{\d \xi^2} +
\left ( -\frac{1}{4} + \frac{n Z}{|\xi|+\xi_m} \right ) \right ] 
Z(\xi) = 0.
\label{eq:Coullike}
\ee
This equation may be solved in terms of Whittaker's functions \cite{Abro70}. 
First, we have
\be
Z_1 (\xi) = A_\pm M_{nZ,1/2} (|\xi| + \xi_m) = 
    A_\pm (|\xi| + \xi_m) \F11 (1-nZ,2,|\xi|+\xi_m )
    e^{-(|\xi|+\xi_m)/2}
\label{eq:Rydberg}
\ee
where $A_\pm$ are the normalization constants for $\xi>0$ and $\xi<0$ 
respectively. Unless $n Z$ is an integer, these solutions tend to infinity 
as $\xi$ goes to infinity.  

As with the equation for an unmagnetized Coulomb potential, there
exists an additional set of solutions.  For the three-dimensional Coulomb
problem, this solution diverges at the origin and is unphysical.
However, here we can obtain a well behaved solution.  By the method of
reduction of order, we obtain the alternative solutions,
\be
Z_2 (\xi) = A_\pm W_{nZ,1/2} (|\xi| + \xi_m) = 
A_\pm (|\xi| + \xi_m) \F11 (1-nZ,2,|\xi|+\xi_m )
e^{-(|\xi|+\xi_m)/2} \int^{|\xi|+\xi_m}
\frac{e^t}{\left (t \F11 (1-nZ,2,t) \right)^2} \d t.
\label{eq:cuspy}
\ee
These solutions agree with earlier treatments of the one-dimensional
hydrogen atom \cite{Loud59,Hain69}.  For integer values of $nZ$, the
integral in \eqref{cuspy} diverges; therefore, the eigenvalues differ
from those of the unmagnetized Coulomb potential.  Additionally for the
unmagnetized Coulomb potential, $\xi_m=0$ and the prefactor of
$(|\xi|+\xi_m)$ is absent.  We find that in this case, this wavefunction
diverges as $\xi^{-1}$ near the origin and only the counterparts of the
states given by \eqref{Rydberg} are physical. 

The solutions to \eqref{Schapprox} will be made of a linear combination
of $Z_1$ and $Z_2$.  For a given magnetic quantum number $m$, the
excitations along the magnetic field axis will be denoted by $\nu$ with
$\nu=0$ being the ground state.  Determining the ground eigenvalue of
\eqref{Coullike} for a given value of $\xi_m$ proceeds in reverse.
Since the ground state is even, we have $A_+=A_-$ and $Z'(0)=0$.  One
first selects a value for $0<nZ<1$.  To have the correct behavior as
$z\rarrow \infty$, we perform the integral of \eqref{cuspy} from
$|\xi|+\xi_m$ to $\infty$ and calculate $Z_{m0}(\xi)$ for $\xi_m=0$.

With the calculated function, one can determine where $Z'_{m0}(\xi)=0$ and
use this as the value of $\xi_m$ corresponding to the eigenvalue $n
Z$.  The value of $\xi_m$ is simply related to the field strength,
\be
\beta = \frac{2 k_m^2}{\xi_m^2} (-\lambda) 
\ee

\subsection{First-order Binding Energies}

As an example we take $Z=1$ and $n=1/\sqrt{15.58}$.  This corresponds
to a bound state ($|000\rangle$) with an energy of 15.58 Ry.  We find
$\xi_0=0.141$ which yields $\beta=1000$.  For $\beta=1000$,
\jcite{Ruder \etal}{Rude94} obtain a binding energy for the $m=0, \nu=0$
state of 18.60986 Ry.  However, it is straightforward to improve upon
our estimate of the binding energy by treating the small difference
between the approximate and effective potential as a perturbation.  We
obtain
\be
E_{m0}^{(1)} = \left < Z_{m0} | H' | Z_{m0} \right >
\ee
where $H'=V_\rmscr{eff}-V_\rmscr{approx}$.  We then obtain the
binding energy to first order of 18.48 Ry for $\beta=1000$.

This technique may also be applied to states with $m<0$ by using the
appropriate value for $k_m$ in \eqref{Schapprox}.  For example, also for
$\beta=1000$ and $m=-1$ ($|0-10\rangle$), we obtain the zeroth order
binding energy of 10.45 Ry and the first-order corrected value of 13.71
Ry compared to the result of \jcite{Ruder \etal}{Rude94} of 13.90394 Ry.
Since \eqref{potapprox} is a better approximation to the effective
potential for electrons in the $m=0$ state than in $m>0$ states we
obtain eigenvalues to first order within 0.7 \% of the fully numerical
treatment for $\beta \ge 1000$ for these states (compared to within 1.4
\% for $m=-1$ states). 

\begin{figure}
\plotonesc{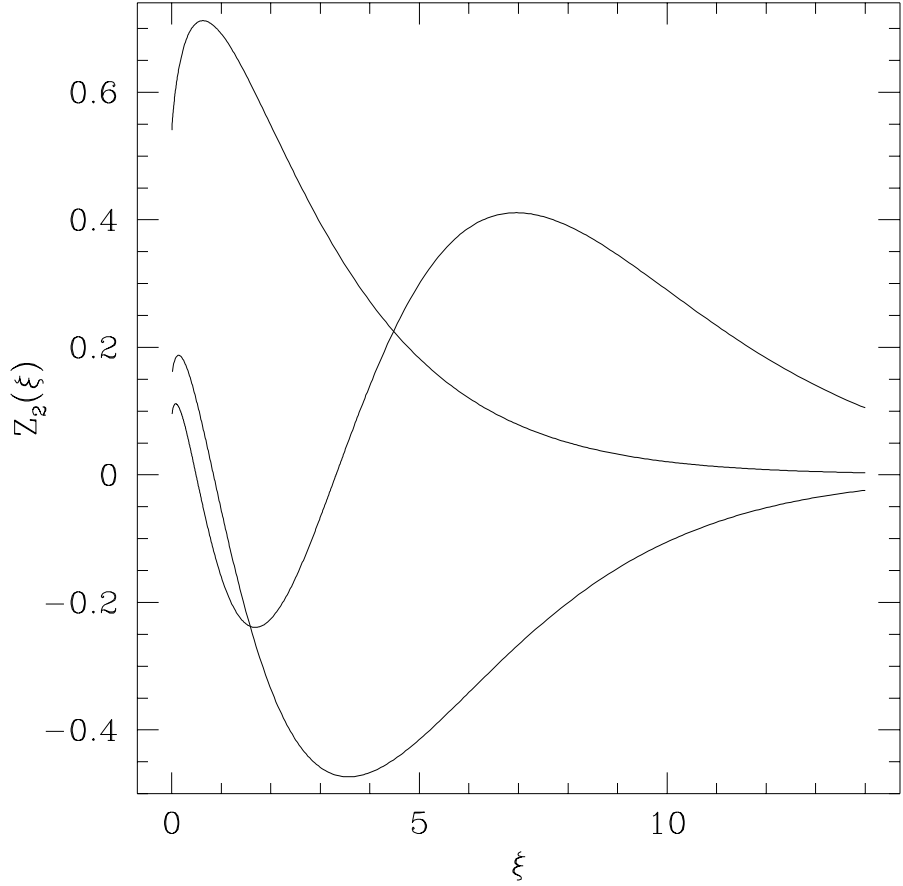}{0.7}
\caption{The function $Z_2(\xi)$ for $\xi_0=0$ for $nZ=1/2, 3/2, 5/2$.}
\label{fig:Z2}
\end{figure}

To calculate the wavefunctions with $\nu>0$, we calculate
$Z_2(\xi)$ for $n Z>1$ and use the first extremum or zero of
$Z_2(\xi)$ as the value of $\xi_0$ for the even and odd solutions
respectively.  \figref{Z2} depicts $Z_2(\xi)$
for several values of $nZ$.  For $nZ$ between $k$ and $k+1$, 
$Z_2(x)$ has $k$ zeros and $k+1$ extrema.  Therefore, we find that the
$\nu>0$ states have zeroth-order binding energies of fractions of a Rydberg.
The calculation of $Z_2(\xi)$ is complicated by the fact that the
function $Z_1(\xi)$ also has zeros in the range of integration from
$\xi$ to $\infty$ which make \eqref{cuspy} ill defined.
To pass over the singularities in the integrand, we integrate
the differential equation \ref{eq:Coullike} directly.

For smaller values of $nZ$ in the range $k$ to $k+1$, the first zeros
and extrema approach $\xi=0$.  Therefore, for larger values of
$\beta$, the zeroth order eigenvalues of the $\nu>0$ spectrum approach
the Bohr energies.  The energies of the odd states approach the Bohr
energies from below (\ie they are more weakly bound), and 
the even states with the same number of nodes are yet more weakly bound
\cite{Hain69}.

Our first-order adiabatic approximation is less accurate for smaller field
strengths.  For $\beta=100$ and $m=0$ ($|000\rangle$), we obtain a first-order corrected
eigenvalue of 9.348 Ry compared to the numerically derived value of
9.4531 Ry (a difference of 1.1 \%).  However, for fields of $B>5\times
10^{11}$ G, the wavefunctions and binding energies derived in this
section for $m<3$ and arbitrary $\nu$ are sufficiently accurate for all
but the most precise analyses. 

\subsection{Perturbed Wavefunctions}

To obtain first order corrections to the wavefunctions
$Z_{m\nu}$ and second order corrections to the binding
energies, we follow the standard techniques for time-independent
perturbation theory \cite{Bran89}.  We must calculate the following
quantities
\be
H'_{\nu\mu} =  \left < Z_{m\nu} | H' | Z_{m\mu} \right >
\ee
for a particular value of $\beta$.  Since both $V_\rmscr{eff}$ and
$V_\rmscr{approx}$ are symmetric about $z=0$, $H'_{\nu\mu}$ is zero for
$\nu$ odd and $\mu$ even.

We obtain 
\be
Z^{(1)}_{m\nu} = \sum_{\mu\neq\nu}
\frac{H'_{\mu\nu}}{E^{(0)}_\nu-E^{(0)}_\mu} Z^{(0)}_{m\mu}
\ee
and
\be
E^{(2)}_{m\nu}  = \sum_{\mu\neq\nu}
\frac{|H'_{\mu\nu}|^2}{E^{(0)}_\nu-E^{(0)}_\mu}
\ee

\begin{figure}
\plottwo{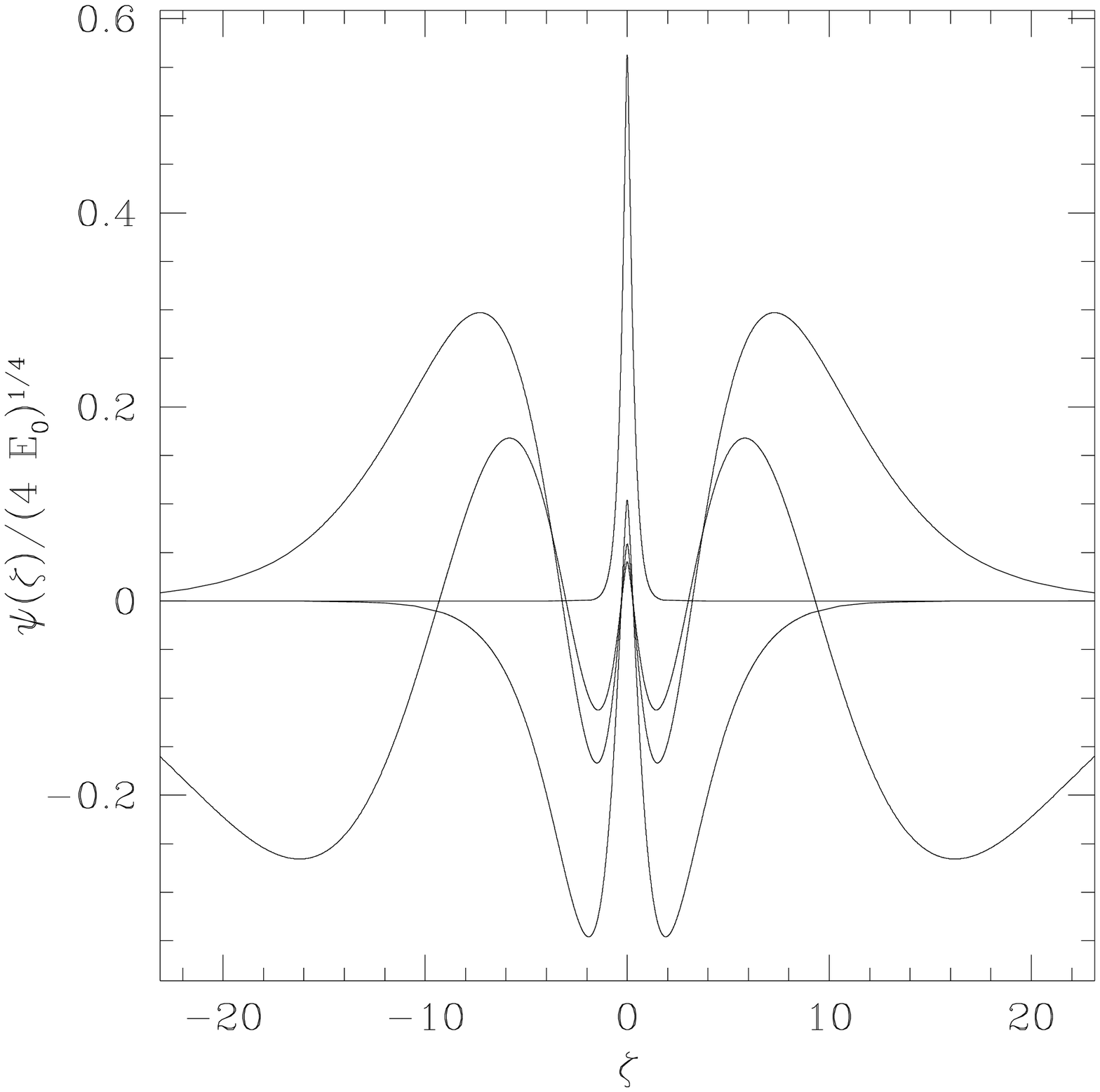}{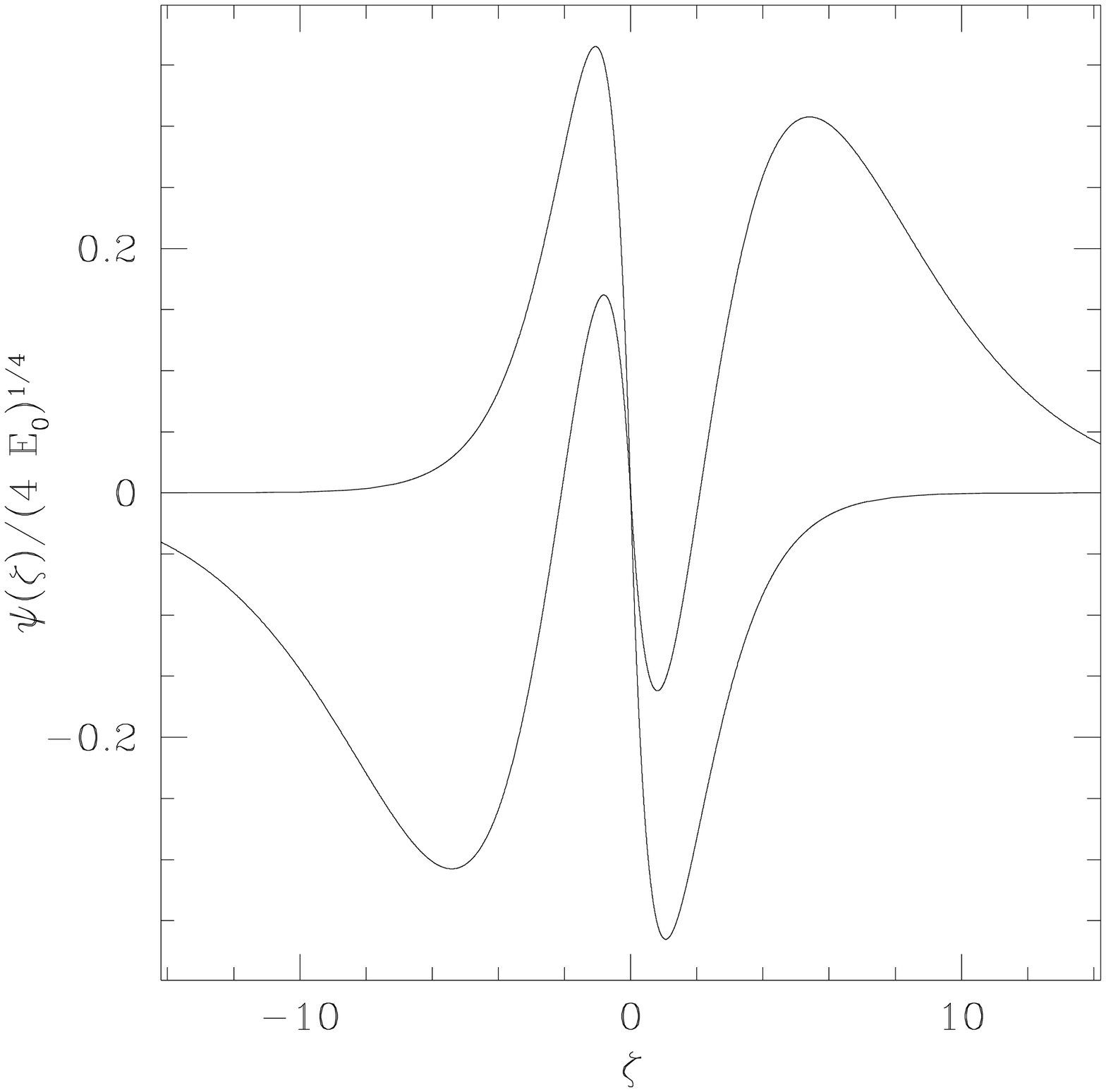}
\caption[The axial wavefunctions of hydrogen in an intense magnetic
field (analytic calculation)]
{The axial wavefunctions of hydrogen in an intense magnetic field
(analytic calculation) for $\beta=1000$.  The left panel depicts the
first four even states with axial excitations ($|000\rangle,
|002\rangle, |004\rangle, |006\rangle$). The right panel shows the first
two odd states ($|001\rangle, |003\rangle$).}
\label{fig:analwf}
\end{figure}

For $\beta=1000$, the mixing among the $\nu$ states is on the order of
a few percent.  The second order corrections to the binding energies
for the ground ($\nu=0$) state is $10^{-3}$ times the first order
correction.  For the excited states with $\nu<6$ the second order
correction is less than six percent of the first-order correction;
we quote the binding energies to first order for the several of the 
most bound levels of hydrogen for $\beta \geq 1000$ in
\tabref{E1} and depict the wavefunctions to zeroth order for
$\beta=1000$ in \figref{analwf}.

\section{Numerical Solution}

\subsection{The Basis Set}
We can make substantial progress by carefully selecting a basis to
expand the solutions $Z_{\nu m}$.  Specifically, we choose
\be
Z_{\mu m}(z) = \sum_{k=0}^\infty A_{\mu m k} {\cal G}_k(z) 
\label{eq:gexpansion}
\ee
where
\def\cGk{{\cal G}_k}
\def\cGl{{\cal G}_l}
\be
\cGk(z) = \frac{1}{(2 \pi)^{1/4} \sqrt{ a_Z 2^k k!} }
H_k \left ( \frac{z}{\sqrt{2} a_Z} \right ) \exp \left (- \frac{z^2}{4
a_Z^2} \right ).
\label{eq:gbasis}
\ee
$H_k(z)$ are the Hermite polynomials which are orthogonal on the
interval $-\infty$ to $\infty$ with the weighting function
$\exp(-z^2)$.  The $\cGk$ are the solutions to the Schr\"odinger equation
for a harmonic oscillator potential; consequently, they provide a
complete set for expanding the functions $Z_{\nu m}(z)$.

To obtain the coefficients in the expansion, we calculate the matrix
\be M_{kl} = \langle \cGk | H_z | \cGl \rangle \ee which is
a function of $a_Z$ and the azimuthal state given by $m$.  We calculate
this matrix for $k,l < N$ ($N=5-50$) and diagonalize it.  The
eigenvalues of this matrix ($\lambda_\nu$) are $E_{\nu m}$, and the
eigenvectors are the coefficients $A_{\nu m k}$ in \eqref{gexpansion}.
Additionally, the functions $Z_{\nu m}(z)$ and $\cGk(z)$ have
definite parity; consequently, for even parity solutions to
\eqref{zSchro}, only the elements of $M_{kl}$ with $k$ and $l$ even need
to be calculated.  This reduces the size of the matrix from $N^2$ to
$N^2/4$. 

Because the number of basis functions used is not infinite, we cannot
expect the expansion to span the Hilbert space of solutions to
\eqref{zSchro}.  To estimate the solution, we vary $a_Z$ to minimize the
eigenvalue $\lambda_\nu$ corresponding to the bound state that we are
interested in.  By using an expansion of the form \eqref{gexpansion},
the binding energies and wavefunctions may be estimated for excited
states along the $z$-axis. 

\begin{figure}
\plottwo{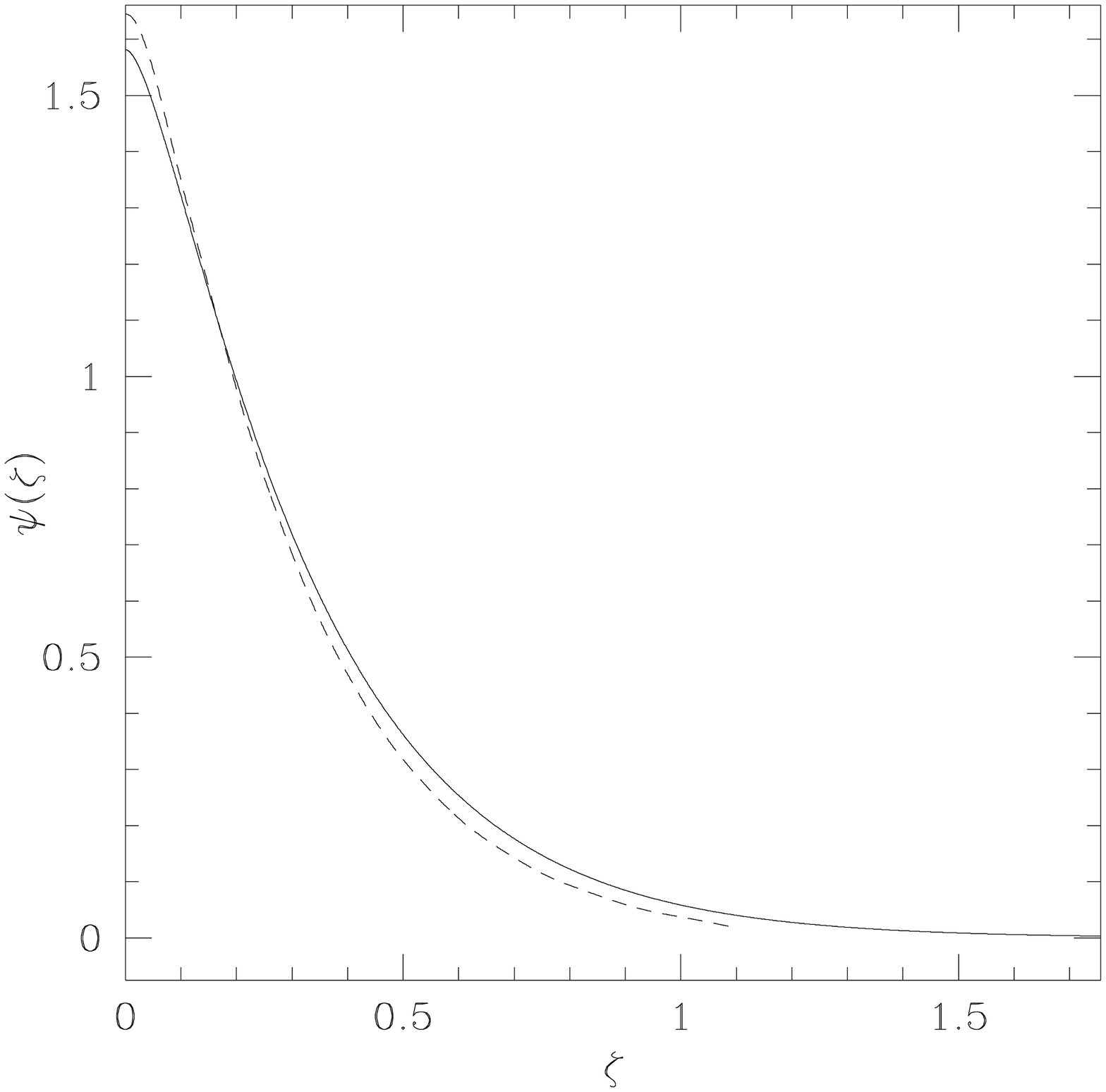}{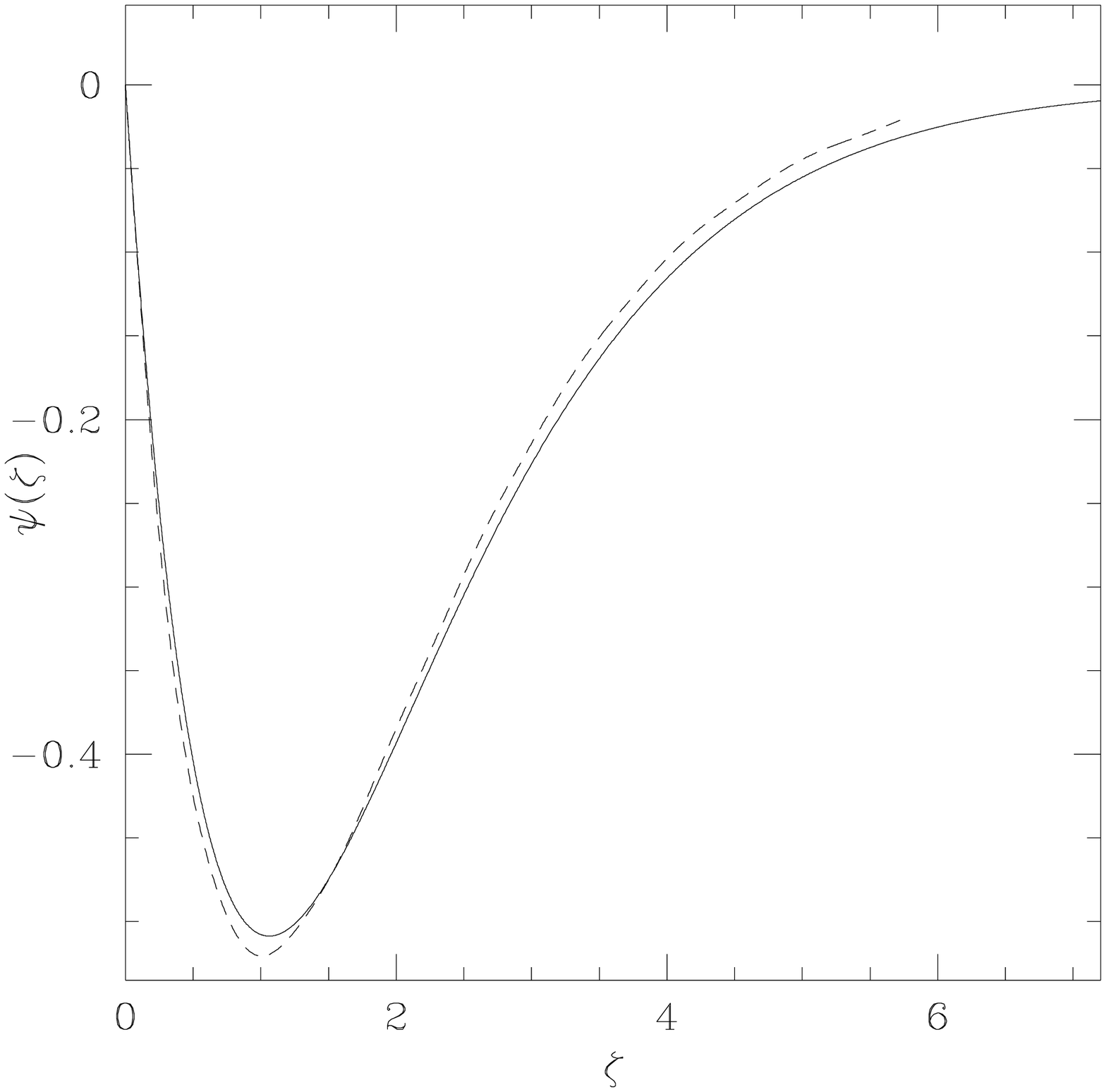}
\caption[{A comparison of numerical and analytic wavefunctions for
hydrogen}]{A comparison of numerical and analytic wavefunctions for
hydrogen.  Both panels are for $\beta=1000$.  The left panel displays
the state $|000\rangle$, and the right shows $|001\rangle$.
The dashed line traces the numerical results with the first
31 $\cGk$.  The solid line traces the zeroth-order analytic solutions.}
\label{fig:boundcmp}
\end{figure}

Although the functions $\cGk$ satisfy a much different equation
from \eqref{zSchro}, if sufficiently many Gauss-Hermite functions are
included, we can obtain highly accurate eigenvalues and eigenvectors.
For the ground state ($|000\rangle$) with the first 31 $\cGk$, we
obtain a binding energy of 18.5579 Ry at $\beta=1000$, within a factor
of $3\times 10^{-3}$ of the result of Ruder \etal, 18.60986 Ry.  The
results are equally accurate for the first excited state ($|001\rangle$);
however, states with more nodes require additional terms in the
expansion to achieve the same accuracy.  \figref{boundcmp} compares the
zeroth-order analytic wavefunction with the numerical wavefunction
determined by this technique. 

Obtaining an additional few parts per thousand in accuracy can only
justify a portion of the additional computation required for this
numerical technique; however, this technique may be applied to solve
the Schr\"odinger equation for potentials more complicated than \eqref{pot}. 

\subsection{The H$_2^+$ molecule}
Before proceeding to the multi-electron problem, we study the binding
energy of the H$_2^+$ molecule in an intense magnetic field.
We restrict our attention to the case where the axis of the molecule is
aligned with the magnetic field direction.  This system retains the
symmetry under parity of hydrogen, so the numerical technique may be
applied directly with only two alterations.

\begin{figure}
\plotonesc{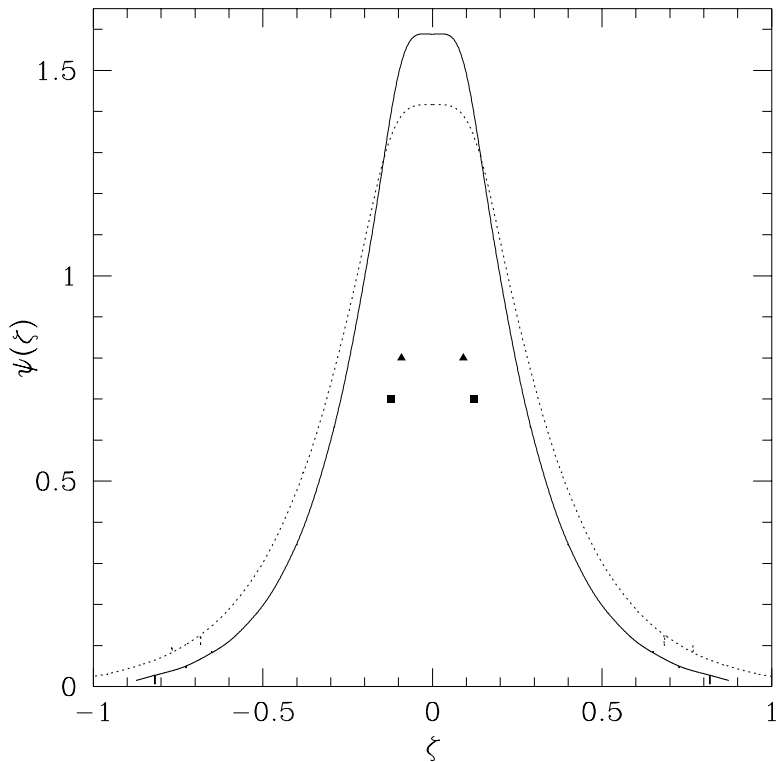}{0.6}
\caption[{The ground and first-excited state of H$_2^+$}]
{The ground and first-excited state of H$_2^+$.  The solid line traces
$|000\rangle$, and the dashed line follows $|0-10\rangle$.  The
triangles give the positions of the protons for the ground state and the
squares for the excited state.}
\label{fig:h2pwf}
\end{figure}

The effective potential is now given by
\be
V_{\rmscr{eff},0m,\rmscr{H}_2^+}(z) = V_{\rmscr{eff},0m}(z+a) + V_{\rmscr{eff},0m}(z-a)
\ee
and we must vary the internuclear separation $2 a$ to find the minimum
binding energy for the entire system (the Born-Oppenheimer
approximation, \eg \cite{Bran89}).  We find the ground state,
$|000\rangle$, at $\beta=1000$ has a binding energy of 28.3457 Ry,
compared to the \jcite{LeGuillot \& Zinn-Justin}{LeGu84} result of 28.362
Ry. The internuclear separation is $0.1818 a_0$;  \jcite{LeGuillot \&
Zinn-Justin}{LeGu84} find $0.181 a_0$.  \figref{h2pwf} depicts the
wavefunctions of the ground and first excited state $|0-10\rangle$ for
H$_2^+$.

The accuracy of our analysis is insufficient to determine if the
ungerade state is slightly bound or unbound relative to a hydrogen atom
plus a proton.  However, in the magnetic case, the electron may be
excited into the $|0m0\rangle$ states whose axial wavefunctions are
similar to that of the $|000\rangle$ state.  The $|0-10\rangle$ state is
much less bound at 20.4252 Ry than the $|000\rangle$ state (compared to
18.5579 Ry for the H+p system).  For stronger fields, the $|0-20\rangle$
and more excited states are bound relative to the H+p system.

\tabref{h2penergy} depicts the numerical results for the ground
and first excited state of H$_2^+$ in an intense magnetic field.  The
ratio of the binding energies of the $|000\rangle$ and $|0-10\rangle$
for H$_2^+$ is approximately equal to the ratio the energies of the
same states of hydrogen and the same magnetic field strength.  This
observation provides a quick way to estimate the energies of the
excited states of H$_2^+$ from the binding energy of the ground state.

\tabref{h2penergyLopez} presents results calculated for different values of 
magnetic field.   Our values differ by less than 0.5 \% for $B\geq 10^{12}$~G and
by $\sim 1 \%$ for the two weaker fields considered.  We see that the first excited
state of H$_2^+$ becomes bound relative to hydrogen atom and a proton at 
$B\approx 10^{12}$~G.  Furthermore, a comparison of 
\tabref{h2penergyLopez} with Table~3 of \jcite{Lopez \etal}{Lope97} shows that 
the ungerade 
state is unbound for $B\geq 10^{11}$~G.

\section{The Multiple Electron Problem}

\subsection{Approach and Results}

To calculate the atomic structure of multi-electron atoms, we employ a
single-configuration Hartree-Fock technique.  Specifically, we assume
that the multi-electron wavefunction is given by a single Slater
determinant of one-electron wavefunctions.  These wavefunctions are
varied to minimize the total energy of the system given the constraint
that each one-electron wavefunction remains normalized.

This minimization results in the following eigenvalue equations for the
individual wavefunctions,
\be
F(1) \psi_i(1) = \epsilon_i \psi_i(1)
\ee
where $1$ denotes the spin and spatial coordinates of the electron
\ie ${\bf r}_1, \sigma_1$.

The operator $F(1)$ is the sum of a kinetic and potential energy term
\be
F(1) = H_0(1) + V(1)
\ee
where the kinetic term is given by the one-particle Schr\"{o}dinger equation
of an electron in the Coulomb field of the nucleus.

The potential energy consists of a direct and exchange interaction with
the other electrons
\be
V(1) = \sum_j \left [ J_j(1)-K_j(1) \right ] 
\ee
where
\ba
J_j(1) \psi_i(1) &=& \left [ \int d\tau_2 \psi_j^*(2) \left (
\frac{e^2}{r_{12}} \right ) \psi_j(2) \right ] \psi_i(1) \\
K_j(1) \psi_i(1) &=& \left [ \int d\tau_2 \psi_j^*(2) \left (
\frac{e^2}{r_{12}} \right ) \psi_i(2) \right ] \psi_j(1) 
\ea
Rather than solve the eigenvalue equations directly, we calculate the
total energy of the system given a set of wavefunctions and minimize
this energy by varying the parameters of the wavefunctions.

In a sufficiently strong magnetic field, these equations for the atomic
structure become approximately separable in cylindrical coordinates. 
With this in mind, we take the trial wavefunctions to be of the form
\be
\psi_i(1) = Z(z) R(\rho,\phi) \chi(\sigma).
\ee
Since we are looking
for the ground state of these atoms, we take all the electron spins
antialigned with the field and the radial wavefunction to be given by
$n=0$ Landau states with each electron occupying a different $m$
state.  We obtain
\be
\psi_i(1) = Z_i(z) R_{0m_i}(\rho,\phi)
\chi_{-\frac{1}{2}} (\sigma)
\ee
where
\be
R_{0m}(\rho,\phi) = \frac{1}{\sqrt{2\pi|m|!}} \frac{1}{a_H}
\exp \left ( -\frac{\rho^2}{4 a_H^2} \right )
 \left ( \frac{\rho}{\sqrt{2}a_H} \right )^{|m|} e^{im\phi}
\ee
and $a_H=\sqrt{\hbar c/|e| B}$.  We suppress the spin portion of the
wavefunction and use the natural length and energy units of the problem
$a_H, e^2/a_H$.

The total energy of the system is given by
\be
E = \sum_i \left < \psi_i(1) | F(1) | \psi_i(1) \right >.
\ee
To expedite the calculation we can integrate over the known
wavefunctions in the $\rho$ and $\phi$ coordinates. Specifically, we
begin with the integral over $\phi$ in the potential energy terms
\ba
\left < \psi_i(1) | J_j(1) | \psi_i(1) \right > &=& e^2
\int \rho_1 d \rho_1 d z_1
Z^*_i(z_1) R^*_{0m_i}(\rho_1) Z_i(z_1) R_{0m_i}(\rho_1) \\*
& & \times
\int \rho_2 d \rho_2 d z_2
Z^*_j(z_2) R^*_{0m_j}(\rho_2) Z_j(z_2) R_{0m_j}(\rho_2)
f(\rho_1,\rho_2,z_1-z_2) 
 \\
\left < \psi_i(1) | K_j(1) | \psi_i(1)  \right > &=& e^2
\int \rho_1 d \rho_1 d z_1
 Z^*_i(z_1) R^*_{0m_i}(\rho_1) Z_j(z_1) R_{0m_j}(\rho_1) \\*
& & \times
\int \rho_2 d \rho_2 d z_2 
 Z^*_j(z_2) R^*_{0m_j}(\rho_2) Z_i(z_2) R_{0m_i}(\rho_2)
g(m_i-m_j,\rho_1,\rho_2,z_1-z_2) 
\ea
where
\ba
f(\rho_1,\rho_2,z_1-z_2) &=&
\int d \phi_1 \int d \phi_2
\frac{1}{\sqrt{\rho_1^2+\rho_2^2+(z_1-z_2)^2-2\rho_1\rho_2\cos(\phi_1-\phi_2)}}
 \\
g(m_i-m_j,\rho_1,\rho_2,z_1-z_2) &=&
\int d \phi_1 \int d \phi_2
\frac{e^{i(m_j-m_i)(\phi_1-\phi_2)}}
{\sqrt{\rho_1^2+\rho_2^2+(z_1-z_2)^2-2\rho_1\rho_2\cos(\phi_1-\phi_2)}}
\ea
The expressions for the functions $f$ and $g$ may be simplified by the
substitution
$\phi_1-\phi_2 = 2 ( \beta + \pi/2)$ and the definition
\be
k^2 = \frac{4 \rho_1 \rho_2 }{(\rho_1+\rho_2)^2 + (z_1-z_2)^2}
\ee
resulting in
\ba
f(\rho_1,\rho_2,z_1-z_2) &=&
\frac{8 \pi}{\sqrt{(\rho_1+\rho_2)^2+(z_1-z_2)^2}}
\int_0^{\pi/2} d \beta \frac{1}{\sqrt{1-k^2\sin^2\beta}} =
\frac{8 \pi}{\sqrt{(\rho_1+\rho_2)^2+(z_1-z_2)^2}}
F \left ( \frac{\pi}{2},k \right ) \\
g(m_i-m_j,\rho_1,\rho_2,z_1-z_2) &=&
\frac{8 \pi}{\sqrt{(\rho_1+\rho_2)^2+(z_1-z_2)^2}}
\int_0^{\pi/2} d \beta \frac{e^{2 i (m_j-m_i)(\beta+\pi/2)}}
{\sqrt{1-k^2 \sin^2\beta}} \\
&=&
\frac{8 \pi}{\sqrt{(\rho_1+\rho_2)^2+(z_1-z_2)^2}}
\int_0^{\pi/2} d \beta \frac{\cos(2 (m_j-m_i)(\beta+\pi/2))}
{\sqrt{1-k^2 \sin^2\beta}}
\ea
where $F \left ( \frac{\pi}{2},k \right )$ is the complete Legendre
elliptic integral of the first kind.  The imaginary portion of the
integral for $g$ must be zero since the Hamiltonian is hermitian (i.e.
unitarity).  This may be seen by expanding the denominator in powers of
$\sin^2\beta$ and multiplying this series by $i \sin(2 (m_i-m_j)
(\beta+\pi/2))$.  The integral of this term is zero.

Numerical Recipes \cite{Pres88} provides routines to efficiently calculate 
$F \left ( \frac{\pi}{2},k \right )=\rmmat{\tt cel}(\sqrt{1-k^2},1,1,1)$ 
(unfortunately, this routine is absent from the latest edition).
Furthermore, for $|m_i-m_j|=1$, we can use the same routine to
calculate $g$,
\be
g(\pm 1,\rho_1,\rho_2,z_1-z_2) =
\frac{8 \pi}{\sqrt{(\rho_1+\rho_2)^2+(z_1-z_2)^2}}
\rmmat{\tt cel}(\sqrt{1-k^2},1,-1,1). 
\ee
For $|m_i-m_j|>1$, we must perform the integral numerically.

Furthermore, we can gain insight on both the functions $f$ and $g$ by
expanding them in the limit of large $\Delta z=|z_i-z_j|$.
\ba
f(\rho_1,\rho_2,z_1-z_2) &=& (2 \pi)^2
\Biggr [ \frac{1}{\Delta z} - \frac{1}{2} \frac{\rho_1^2 + \rho_2^2}{\Delta z}
 + \frac{3}{8} \frac{ \rho_1^4 + 4 \rho_1^2 \rho_2^2 + \rho_2^4 }{\Delta z^5}
 -\frac{5}{16}
\frac{\rho_1^6 + 9 \left ( \rho_1^4 \rho_2^2 + \rho_1^2 \rho_2^4 \right
) + \rho_2^6 }{\Delta z^7} \nonumber \\*
& & ~~~ + \frac{35}{128} \frac{ \rho_1^8 + 16 \left ( \rho_1^6 \rho_2^2 +
\rho_1^2 \rho_2^6 \right ) + 36 \rho_1^4 \rho_2^4 + \rho_2^8}{\Delta
z^9} + {\cal O} \left ( \frac{1}{\Delta z^{11}} \right ) \Biggr ] \\
g(0,\rho_1,\rho_2,z_1-z_2) &=& f(0,\rho_1,\rho_2,z_1-z_2) \\
g(\pm 1,\rho_1,\rho_2,z_1-z_2) &=& (2 \pi)^2 \Biggr [
\frac{1}{2} \frac{\rho_1 \rho_2}{\Delta z^3}
- \frac{3}{4} \frac{\rho_1 \rho_2^3 + \rho_1^3 \rho_2}{\Delta z^5}
+ \frac{15}{16} \frac{ \rho_1 \rho_2^5 + 3 \rho_1^3 \rho_2^3 + \rho_1^5
\rho_2 } {\Delta z^7} \nonumber \\*
& & ~~~
- \frac{35}{32} \frac{\rho_1 \rho_2^7 +
6 \left (\rho_1^5 \rho_2^3 + \rho_1^3 \rho_2^5 \right ) + \rho_1^7
\rho_2}{\Delta z^9}
+ {\cal O} \left ( \frac{1}{\Delta z^{11}} \right ) \Biggr ] \\
g(\pm 2,\rho_1,\rho_2,z_1-z_2) &=& (2 \pi)^2 \Biggr [
\frac{3}{8} \frac{\rho_1^2 \rho_2^2}{\Delta z^5}
- \frac{15}{16} \frac{\rho_1^2 \rho_2^4 + \rho_1^4 \rho_2^2}{\Delta z^7}
+ \frac{105}{64} \frac{\rho_1^6 \rho_2^2 \frac{8}{3} \rho_1^4 \rho_2^4 +
\rho_1^2 \rho_2^6 }{\Delta z^9} \nonumber \\*
& & ~~~
- \frac{315}{128} \frac{ \rho_1^2 \rho_2^8 + 
5 \left ( \rho_1^4 \rho_2^6 + \rho_1^6 \rho_2^4 \right )
+ \rho_1^8 \rho_2^2 }{\Delta z^{11}}
+ {\cal O} \left ( \frac{1}{\Delta z^{13}} \right ) \Biggr ] \\
g(\pm 3,\rho_1,\rho_2,z_1-z_2) &=& (2 \pi)^2 \Biggr [
\frac{5}{16} \frac{\rho_1^3 \rho_2^3}{\Delta z^7} 
- \frac{35}{32} \frac{\rho_1^5 \rho_2^3 + \rho_1^3 \rho_2^5}{\Delta z^9}
+ {\cal O} \left ( \frac{1}{\Delta z^{11}} \right ) \Biggr ] \\
g(\pm 4,\rho_1,\rho_2,z_1-z_2) &=& (2 \pi)^2 \Biggr [
\frac{35}{128} \frac{\rho_1^4 \rho_2^4}{\Delta z^9}
+ {\cal O} \left ( \frac{1}{\Delta z^{11}} \right ) \Biggr ] 
\ea
and in general
\be
g(\pm \Delta m,\rho_1,\rho_2,z_1-z_2) \propto
\frac{\left (\rho_1\rho_2\right)^{\Delta m}}{\Delta z^{2\Delta m +1}}
\ee
to leading order in $1/\Delta z$.

In the limit of large $\Delta z$, the integrals over the radial
wavefunctions may evaluated using these expansions.  This calculation is
speeded by the observation that
\be
\int 2 \pi \rho d \rho R_{0m_1} (\rho) R_{0m_2} (\rho) \rho^n
= \sqrt{ \frac{2^n}{|m_1|!|m_2|!} } \Gamma \left (
\frac{|m_1|+|m_2|+n}{2} + 1 \right )
\ee
which may be proven by using the normalization condition of the
functions $R_{0m}(\rho)$ and analytically continuing the factorial
function with the Gamma function.  For $\Delta z<10$ we have numerically
integrated the functions $f$ and $g$ over the various pairs of Landau states.

After the integration over the radial and angular coordinates, the
energy may now be written as expectation values of operators acting on
the $Z(z)$ wavefunction.  Since each electron is assumed to be in a
particular Landau $m$ level, we can calculate an effective potential
energy between the electron and the nucleus by integrating over
$\rho,\phi$.  The potential is given by \eqref{Veff}. 

The calculational strategy is similar to the single electron case.  The
quantum numbers $\nu, m$ for each electron are chosen ahead of time, and
the wavefunction $Z(z)$ is expanded as \eqref{gexpansion} with each
electron having is own variable value of $a_Z$.  For each
electron $i$, the matrix 
\be
(M_i)_{kl} =  \langle \cGk | F(i) | \cGl \rangle 
\ee
is calculated.

The added complication is that the diagonalization of the matrices $M_i$
must proceed iteratively.  For the given values of $a_Z$, the matrices
are first calculated assuming that the other electrons ($j\neq i$) have
$A_k=1$ for $k=\nu_j$.  Then each electron's matrix is diagonalized and
the $\nu_i$th eigenvector is used to calculate the interelectron
potential for the next iteration.  The matrices converge after $\sim
5-10$ iterations.  Next, the values of $a_Z$ for each electron are
varied to minimize the total energy of the configuration. 

For brevity, we discuss the ground state energies and wavefunctions
for H$_2$, He and HHe as a function of field strength for $\beta\geq
1000$.
For the molecules we again take the molecular axis to be
aligned with the magnetic field direction.  Since we are interested in
the ground states of these species we set $\nu=0$ for all the
electrons and assign the electrons consecutive $m$ quantum numbers
beginning with $m=0$.  Because none of the electrons have axial
excitations, we are interested in only the most negative eigenvalue of
the electron matrices.  This eigenvalue is more efficiently determined
by varying the coefficients in \eqref{gexpansion} directly than by
diagonalizing the electron matrices iteratively.

\tabref{h2energy} gives the binding energies of the most tightly
bound states of H$_2$, He, HHe and H calculated numerically using the
variational method.  The energies for H are within 1.1 \% of the values
quoted by \jcite{Ruder \etal}{Rude94} for weakest field strength common
between the two studies.  For the stronger fields, the agreement is even
closer.  For He the energies are within 2.5 \% of the values of Ruder
\etal for the fields that overlap.

We also computed the binding energies of H$_2$ and H$^-$ and compared
the results with the values found by \jcite{Lai \etal}{Lai92}.  We
interpolated the Lai \etal\ results using a cubic spline with $\ln
\beta$ as the independent variable.  The binding energies for H$_2$ were
within $0.6 - 3$\% of the Lai \etal\ results.  The agreement for
H$^-$ was poorer ranging from $2-7$\%.  The results for H and H$_2^+$
(\tabref{h2penergy}) agree to within 0.9\% of the Lai \etal\ values.

We compare these interpolated results with the results for the species
He, HHe and HHe$^+$ to find that the reaction
\be
\rmmat{H} + \rmmat{He} \rightarrow \rmmat{HHe}
\ee
is exothermic over the range of field strengths considered.  However,
if there is sufficient hydrogen present, the species HHe would quickly be
consumed by the exothermic reaction
\be
\rmmat{H} + \rmmat{HHe} \rightarrow \rmmat{He} + \rmmat{H}_2 
\ee
for these field strengths, unless
\be
\rmmat{H} + \rmmat{HHe} \rightarrow \rmmat{H}_2\rmmat{He} 
\ee
is sufficiently exothermic (the binding energy and configuration of
H$_2$He is beyond the scope of this paper).
The potential production channels for HHe$^+$,
\ba
\rmmat{He} + \rmmat{H}_2 &\rightarrow & \rmmat{HHe}^+ + \rmmat{H}^-, \\
\rmmat{He} + \rmmat{H}_2^+ &\rightarrow & \rmmat{HHe}^+ + \rmmat{H} \rmmat{~and~} \\
\rmmat{He} + \rmmat{H} &\rightarrow & \rmmat{HHe}^+ + \rmmat{e}^- 
\ea
are endothermic over the range of field strengths considered.
We therefore conclude that at least for hydrogen and helium, atoms in
an intense magnetic field are far more cohesive than adhesive.

\subsection{Validity of the Born-Oppenheimer Approximation}

When studying molecules in a intense magnetic field, we have assumed
that the nuclear motion and the electronic motion decouple, \ie the
Born-Oppenheimer approximation.  Schmelcher \etal \cite{Schm88} have
examined the validity of this approximation in the presence of a
strong magnetic field.  They performed a pseudoseparation of the
equations of motion and derive electronic and nuclear equations of
motion.  Because we have assumed throughout that the nuclei are
infinitely massive and that the molecules are aligned with the
magnetic field, the corrections to the Born-Oppenheimer approximation
may be neglected.  However, the techniques outlined here, specifically
the use of one-dimensional Coulomb wavefunctions and Gauss-Hermite
functions as a convenient and compact basis for the electronic
wavefunctions of atoms and molecules in intense magnetic fields, can
be extended to the more general case where these restrictions have
been relaxed.

\section{Conclusions}

We have developed both an analytic and a convenient numerical
technique to accurately calculate the properties of simple atoms and
molecules in an intense magnetic field.  The calculations presented
here complement the earlier work.  We examine two compounds (HHe and
HHe$^+$) in addition to the species studied earlier which may form in
the intensely magnetized outer layers of a neutron star.
Additionally, our technique finds both tightly bound and excited
states efficiently and accurately which is necessary to calculate the
radiative transfer of the neutron star atmosphere.

The techniques presented in this paper complement the recent work in
this area.  They provide moderately high precision with little
computational or algebraic expense.  Most recent work has focussed on
extremely high precision by using a Hartree-Fock-like method to reduce
the three-dimensional problem to three coupled one-dimensional
problems.  Generally, two of the one-dimensional equations are solved
over an functional basis, \ie Legendre polynomials \cite{Mele93} or
spherical harmonics \cite{Jone96}, and the radial differential
equation is solved numerically over a pointlike basis.
\jcite{Fassbinder \etal}{Fass96} and Shertzer \etal
\cite{Sher89,Sher90} use a finite element method throughout.

The spirit of the work presented here is different.  We have solved
for the wavefunctions using a basis for all three coordinates: the
Landau wavefunctions in the angular and radial direction and the
one-dimensional Coulomb wavefunctions or the Gauss-Hermite functions
along the axis of the magnetic field.  The power of this technique is
that the basis functions resemble the actual wavefunctions and
preserve the symmetries of the potential; consequently only a few
basis functions ($\sim 1-2$ and $\sim 20$ respectively) are require to
reach moderately high precision.

The work of Kravchenko \etal \cite{Krav96a,Krav96b} takes an
orthogonal approach and achieves very high precision by solving the
general problem of a hydrogen atom in an arbitrarily strong magnetic
field with a double power series in $\sin \theta$ and $r$.  It remains
to be seen whether this simple and accurate technique can be applied
to more general problems.

The properties of the lowest density layers of a neutron star's crust
determine the spectral characteristics of the radiation expected from
the star.  One possibility is that linear chains of atoms form in the
surface layers \cite{Rude74,Chen74,Flow77,Mull84,Neuh87,Lai92}, and
the atmosphere condenses at finite density. We find that the reactions
between hydrogen and helium are unlikely to affect the formation of
hydrogen or helium chains unless the formation of hydrogen-helium
hybrid chains is favored.

If the envelope is truncated at sufficiently high density, the thermal
isolation of the core can be substantially reduced \cite{Hern85}.
Furthermore, the composition of the outermost layers determines the
spectra from the neutron star (\eg
\cite{Pavl94,Pavl96,Zavl96,Raja97}).  Without understanding magnetized
chemistry in neutron-star atmospheres, is difficult to interpret
observations of these objects.


\clearpage

\begin{table}
\begin{center}
\caption[The zeroth and first-order binding energies of hydrogen in an intense
magnetic field]
{The zeroth and first-order binding energies of hydrogen in an intense
magnetic field in Rydberg units}
\label{tab:E1}
\begin{tabular}{r|rr|rr|rr|rr|rr}
 &
\multicolumn{2}{c|}{$|000\rangle$} &
\multicolumn{2}{c|}{$|0-10\rangle$} &
\multicolumn{2}{c|}{$|0-20\rangle$} &
\multicolumn{2}{c|}{$|001\rangle$} &
\multicolumn{2}{c}{$|002\rangle$} \\
\multicolumn{1}{c|}{$\beta$} &
\multicolumn{1}{c}{$E_0$} &
\multicolumn{1}{c|}{$E_1$} &
\multicolumn{1}{c}{$E_0$} &
\multicolumn{1}{c|}{$E_1$} &
\multicolumn{1}{c}{$E_0$} &
\multicolumn{1}{c|}{$E_1$} &
\multicolumn{1}{c}{$E_0$} &
\multicolumn{1}{c|}{$E_1$} &
\multicolumn{1}{c}{$E_0$} &
\multicolumn{1}{c}{$E_1$} \\ \hline
$1 \times 10^3$ & 15.58 & 18.48 & 10.45 & 13.71 & 8.779 & 11.76 & 0.9401 & 0.9888 & 0.5841 & 0.6215 \\
$2 \times 10^3$ & 18.80 & 22.26 & 12.81 & 16.73 & 10.83 & 14.46 & 0.9559 & 0.9935 & 0.6062 & 0.6322 \\
$5 \times 10^3$ & 23.81 & 28.09 & 16.57 & 21.51 & 14.12 & 18.73 & 0.9710 & 0.9970 & 0.6329 & 0.6560 \\
$1 \times 10^4$ & 28.22 & 33.19 & 19.94 & 25.73 & 17.10 & 22.55 & 0.9790 & 0.9983 & 0.6518 & 0.6730 \\
$2 \times 10^4$ & 33.21 & 38.91 & 23.81 & 30.53 & 20.56 & 26.93 & 0.9849 & 0.9990 & 0.6684 & 0.6880 \\
$5 \times 10^4$ & 40.75 & 47.49 & 29.76 & 37.81 & 25.91 & 33.61 & 0.9903 & 0.9996 & 0.6885 & 0.7060 \\
$1 \times 10^5$ & 47.20 & 54.76 & 34.95 & 44.08 & 30.60 & 39.40 & 0.9931 & 0.9998 & 0.7027 & 0.7188 
\end{tabular}
\end{center}
\end{table}

\begin{table}
\begin{center}
\caption[The binding energy of H$_2^+$ in an intense magnetic field]
{The binding energy of H$_2^+$ in an intense magnetic field.  The
values have been derived numerically and the final column gives the
numerically derived binding energy of the ground state of H for
comparison.}
\label{tab:h2penergy}
\begin{tabular}{r|rr|r}
&
\multicolumn{2}{c|}{H$_2^+$} &
\multicolumn{1}{c}{H} \\
\multicolumn{1}{c|}{$\beta$} &
\multicolumn{1}{c}{$|000\rangle$} &
\multicolumn{1}{c|}{$|0-10\rangle$} &
\multicolumn{1}{c}{$|000\rangle$} \\ \hline 
$1 \times 10^3$ & 28.35 & 20.43 & 18.57 \\
$2 \times 10^3$ & 35.04 & 25.63 & 22.37 \\
$5 \times 10^3$ & 45.77 & 34.08 & 28.25 \\
$1 \times 10^4$ & 55.37 & 41.83 & 33.37 \\
$2 \times 10^4$ & 66.56 & 50.86 & 39.11 \\
$5 \times 10^4$ & 83.45 & 64.95 & 47.70 \\
$1 \times 10^5$ & 98.27 & 77.38 & 54.96
\end{tabular}
\end{center}
\end{table}

\begin{table}
\begin{center}
\caption[The binding energy of H$_2^+$ in an intense magnetic field (comparison)]
{The binding energy of H$_2^+$ in an intense magnetic field for comparison with
the results of \jcite{Lopez \etal}{Lope97}.}
\label{tab:h2penergyLopez}
\begin{tabular}{r|rr|r}
&
\multicolumn{2}{c|}{H$_2^+$} &
\multicolumn{1}{c}{H} \\
\multicolumn{1}{c|}{B (G)} &
\multicolumn{1}{c}{$|000\rangle$} &
\multicolumn{1}{c|}{$|0-10\rangle$} &
\multicolumn{1}{c}{$|000\rangle$} \\ \hline 
$1 \times 10^{11}$ & 7.347 & 4.880 & 5.611 \\
$5 \times 10^{11}$ & 13.37 & 9.188 & 9.568 \\
$1 \times 10^{12}$ & 17.05 & 11.88 & 11.87 \\
$2 \times 10^{12}$ & 21.53 & 15.25 & 14.58 \\
$5 \times 10^{12}$ & 28.89 & 20.85 & 18.89 \\
$1 \times 10^{13}$ & 35.69 & 26.14 & 22.74 
\end{tabular}
\end{center}
\end{table}

\begin{table}
\begin{center}
\caption[The binding energy of He, HHe$^+$ and HHe in an intense magnetic field]
{The binding energy of He, HHe$^+$ and HHe in an intense magnetic field.  The number 
in parenthesis gives the number of free parameters in each variational model.
The electrons occupy the most tightly bound states, $|0m0\rangle$, \eg $|000\rangle$, 
$|0-10\rangle$ and $|0-20\rangle$ for HHe. The values have been derived numerically 
and the final column gives the numerically derived binding energy of the ground state
of H for comparison.}
\label{tab:h2energy}
\begin{tabular}{r|rrr|r}
\multicolumn{1}{c|}{$\beta$} &
\multicolumn{1}{c}{He (6)} &
\multicolumn{1}{c}{HHe$^+$ (5)} &
\multicolumn{1}{c|}{HHe (7)} &
\multicolumn{1}{c}{H (25)}
 \\ \hline 
$1 \times 10^2$ & 32.47 & 35.75 & 42.59 & 9.383 \\
$2 \times 10^2$ & 40.98 & 45.95 & 54.25 & 11.64 \\
$5 \times 10^2$ & 54.95 & 63.07 & 73.36 & 15.28 \\
$1 \times 10^3$ & 67.85 & 79.24 & 90.89 & 18.57 \\
$2 \times 10^3$ & 83.00 & 98.55 & 111.3 & 22.37 \\
$5 \times 10^3$ & 106.9 & 129.7 & 143.1 & 28.25 \\
$1 \times 10^4$ & 127.8 & 157.5 & 168.0 & 33.37 \\
$2 \times 10^4$ & 151.6 & 189.7 & 193.1 & 39.11 \\
$5 \times 10^4$ & 187.6 & 239.6 &       & 47.70 \\
$1 \times 10^5$ & 218.4 & 282.5 &       & 54.96 
\end{tabular}
\end{center}
\end{table}

\clearpage

\end{document}